\documentclass[twocolumn,aps,prb]{revtex4}
\usepackage{amsmath,amssymb,mathrsfs,bm}
\usepackage{graphicx,dcolumn,times}
\begin{document}
\title{Efficient quantum memory for single photon polarization qubits}
\author{Yunfei Wang$^1$\footnotemark[2], Jianfeng Li$^1$\footnotemark[2],  Shanchao Zhang$^1$\footnote[2]{These authors contributed equally to this work.}, Keyu Su$^1$, Yiru Zhou$^1$, Kaiyu Liao$^1$, Shengwang Du$^{1,2}$, Hui Yan$^1$\footnotemark[1] and Shi-Liang Zhu$^{3,1}$\footnote[1]
{email: yanhui@scnu.edu.cn; slzhu@nju.edu.cn}}
\affiliation{
$^1$Guangdong Provincial Key Laboratory of Quantum Engineering and Quantum Materials, SPTE, South China Normal University, Guangzhou 510006, China\\
$^2$Department of Physics, The Hong Kong University of Science and Technology, Clear Water Bay, Kowloon, Hong Kong S.A.R., China\\
$^3$National Laboratory of Solid State Microstructures, School of Physics, Nanjing University, Nanjing 210093, China}

\begin{abstract}
\textbf{A quantum memory, for storing and retrieving flying photonic quantum states, is a key interface for realizing long-distance quantum communication and large-scale quantum computation. While many experimental schemes of high storage-retrieval efficiency have been performed with weak coherent light pulses, all quantum memories for true single photons achieved so far have efficiencies far below 50\%, a threshold value for practical applications. Here, we report the demonstration of a quantum memory for single-photon polarization qubits with an efficiency of $>$85\% and a fidelity of $>$99\%, basing on balanced two-channel electromagnetically induced transparency in laser-cooled rubidium atoms. For the single-channel quantum memory, the optimized efficiency for storing and retrieving single-photon temporal waveforms can be as high as 90.6\%. Our result pushes the photonic quantum memory closer to its practical applications in quantum information processing.}
\end{abstract}
\maketitle

Quantum memories that  allow for storing unknown quantum states and later retrieving them on demand in controllable time delays are essential for realizing repeater-based long-distance quantum networks and large-scale quantum computers \cite{KimbleNature2008, Kuzmich2005, QRep2011, LQCRev, DLCZ}. Unlike classical memories, the efficiency of a quantum memory plays a more important role in preserving the quantum information. For example, beating the quantum no-cloning limit without postselction requires a memory efficiency of more than 50\% \cite{noclonelimit}, which is also necessary for some error correction protocols in one-way quantum computation \cite{errorcorrection}. However, realizing a quantum memory with near-unity values of both memory efficiency and quantum-state fidelity remains a challenge.

A photonic memory is built  on the light-matter interface between flying photonic modes and long-lived matter states.  Efficient optical storages have been demonstrated in various promising schemes, such as 87\% efficiency with photon echo technique \cite{SE69GEM2010, SE87GEM2016}, 30\% with off-resonance Raman interaction\cite{Ram2010Walmsley, Ram2011Walmsley}, and 92\% with electromagnetically induced transparency (EIT)\cite{EIT2014Buchler, Yu2013, LauratNC2018, EIT2018Yu}.
However all these experimental demonstrations have been performed with attenured weak coherent laser pulses.

Although a weak coherent laser pulse can contain an average photon number in or below the single-photon level, the photon number uncertainty makes it not faverable for many protocols of quantum information processing. For example, linear optics quantum computation requires manipulation of single-photon qubits and cannot work with coherent light pulses \cite{KnillNature2001}. It is not obvious that the above demonstrated efficient coherent-pulse optical memory can directly work for single photon Fock states. There are mainly two challenges on the way toward efficient single-photon quantum memories. The first challenge is suppressing the noise. Uncorrelated noise may not affect much on the characterization of coherent pulses in both time and frequency domains, but it degrades significantly the purity of a single photon state. In those experiments with coherent light pulses, because the strong control laser beams are nearly collinear with the memory optical modes, their scattering and coupling to the single-photon detection is extremely difficult to be filtered away for single photon storage. The second challenge is having single photons with controllable spectral-temporal states matching the memory modes, while this is not difficult at all for coherent light pulse shaping. As a result, the highest efficiency records are 49\% for the single-photon temporal waveform \cite{EIT2012Du}, 10\% for single-photon polarization qubit \cite{EIT2013Ding}, and 26.7\% for entanglement \cite{Entanglement, Ram2015Guo}. Boosting up the memory efficiency is crucial to make the quantum memory practical \cite{QRep2011,SE87GEM2016,EIT2018Yu,LauratNC2018,QMem2016Pan}. For example, a 1\% increase in the memory efficiency would increase the entanglement distribution rate by a value beyond 10\% in the repeater-based quantum network\cite{QRep2011,QMem2016Pan}.

\begin{figure*}[ptb]
\begin{center}
\includegraphics[width=18cm]{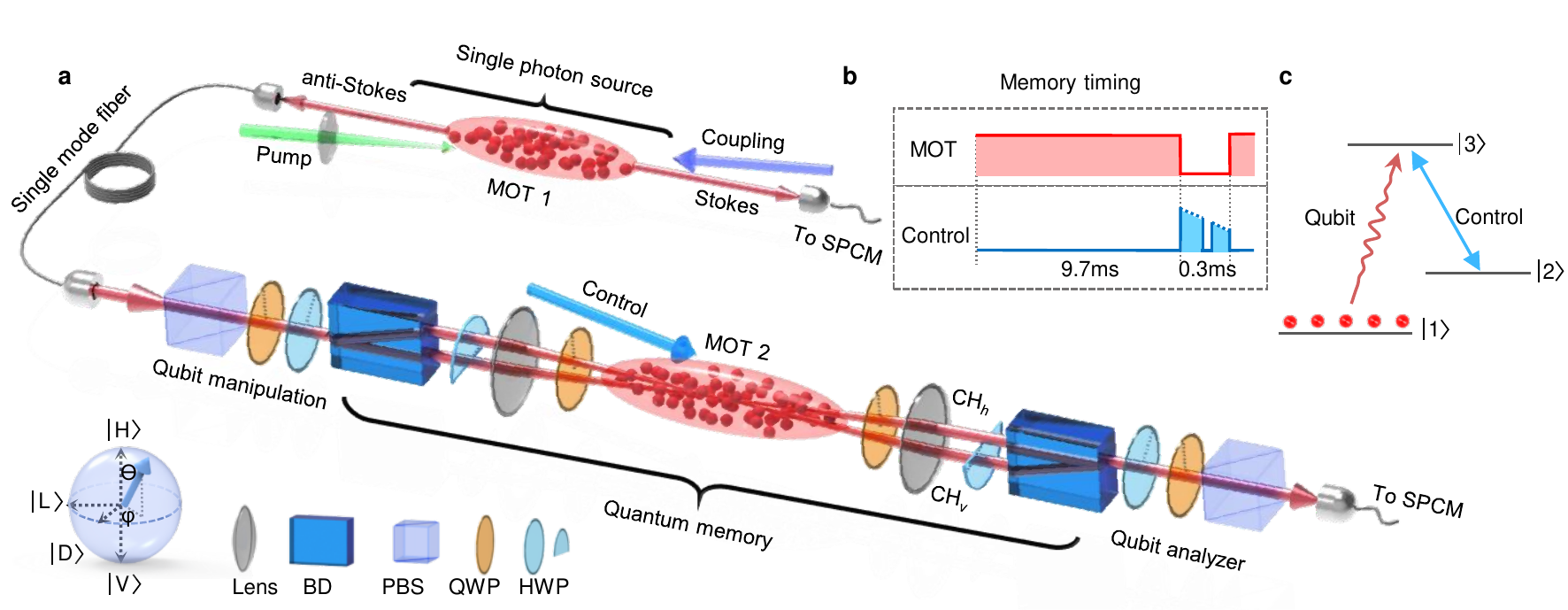}
\caption{\label{fig:sys} \textbf{Experimental setup and energy level scheme of the single-photon quantum memory}.
\textbf{a.} Schematics of experimental optical setup. The cold atoms in the first magneto-optical trap (MOT$_1$) serve as a nonlinear optical medium for producing time-frequency entangled photon pairs, while the cold atoms in the second magneto-optical trap (MOT$_2$) are the medium for the quantum memory. The anti-Stokes photon is coded with an arbitrary polarization state through the qubit manipulation unit (QMU) consisting of a quart-wave plate (QWP) and half-wave plate (HWP). After the QMU, the two orthogonal linear polarizations are separated into two beams by a polarization beam displacer (BD) which are coupled into the two balanced spatial channels CH$_h$ and CH$_v$ of the quantum memory. The memory read outs are recombined at the second BD and the polarization state is measured by the qubit analyzer.
\textbf{b.} The memory operation timing shows the MOT sequence and the optimized control laser intensity time-varying profile in each experimental cycle.
\textbf{c.} The atomic energy level scheme of the quantum memory based on electromagnetically induced transparency (EIT).}
\end{center}
\end{figure*}

Here,  we demonstrate an efficient EIT-based cold-atom quantum memory for single-photon polarization qubits with an efficiency of more than 85\% and a fidelity beyond 99\%. We achieve this by mapping an unknown photonic qubit state in two-dimensional Hilbert space onto two balanced spatial modes in the atomic medium and using a single control laser beam to dynamically write and read the stored qubit. To overcome the first challenge discussed in the above, we set the control laser beam with an angle of $2.2^{o}$ to suppress the scattering and coupling of the strong control laser beam to the qubit modes. Together with other filtering techniques we minimize the system noise for achieving high quantum state fidelity and single-photon purity. Meanwhile, the phase mismatching caused by the large angle separation significantly suppresses the four-wave mixing (FWM) nonlinear process. Secondly, we implement heralded narrowband single photons with controllable temporal waveforms which match the spectral-temporal modes of the memory. We further demonstrate the optimized efficiency of a single-channel temporal mode quantum memory can be as high as 90.6\%. The delay-bandwidth product of the memory is about 10, in which a single-photon qubit can be stored for up to 10 pulse time delay with more than 80\% efficiency below the two-photon threshold. As the EIT-based quantum memory has been demonstrated to be compatible with various quantum schemes \cite{EIT2007Novikova,EIT2008Squeez,EIT2011Pan,EIT2013Ding, QMemRev2009, QMemRev2016}, our work would thus bring it closer to practical applications.

\begin{figure*}
\begin{center}
\includegraphics[width=18cm]{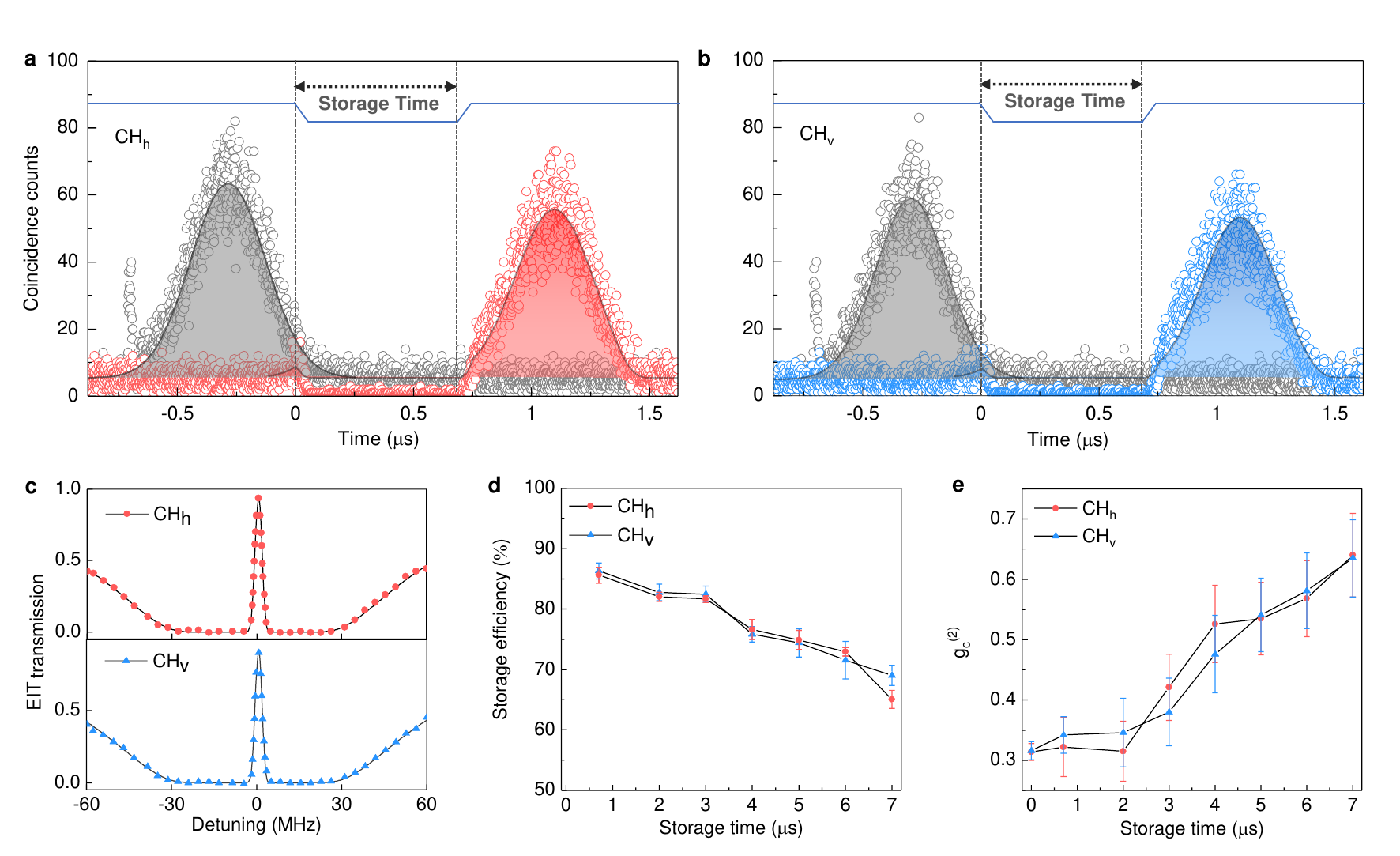}
\caption{\label{fig:wfm2}\textbf{Characterization of the two balanced channels of the quantum memory.}
\textbf{a,b} The temporal waveforms of the heralded single photons before and after one-pulse-delay storage for channels CH$_h$ and CH$_v$. The photon coincidence counts are recorded for 900~s. The solid lines are the best fitted theoretical waveforms. The input waveform is fitted with a Gaussian function while the reterived waveform is numerically calculated according to the theoretical model presented in the Methods.
\textbf{c.} The measured EIT transmission profiles for the quantum memory. The solid lines are the theoretical curves.
\textbf{d.} The storage efficiencies of the channels CH$_h$ and CH$_v$ as functions of the storage time delay.
\textbf{e.} The conditional 2nd-order auto-correlation $g_{c}^{(2)}$ of the retrieved single photons as a function of the storage time delay.  As a reference,  $g_{c}^{(2)}$ of the input single photon is presented as the storage time of 0 ns in the same figures. $g_{c}^{(2)}$ is measured over a coincidence time window of 700 ns. The error bars are from the standard deviation of the statistics.}
\end{center}
\end{figure*}

\begin{figure*}
\begin{center}
\includegraphics[width=18cm]{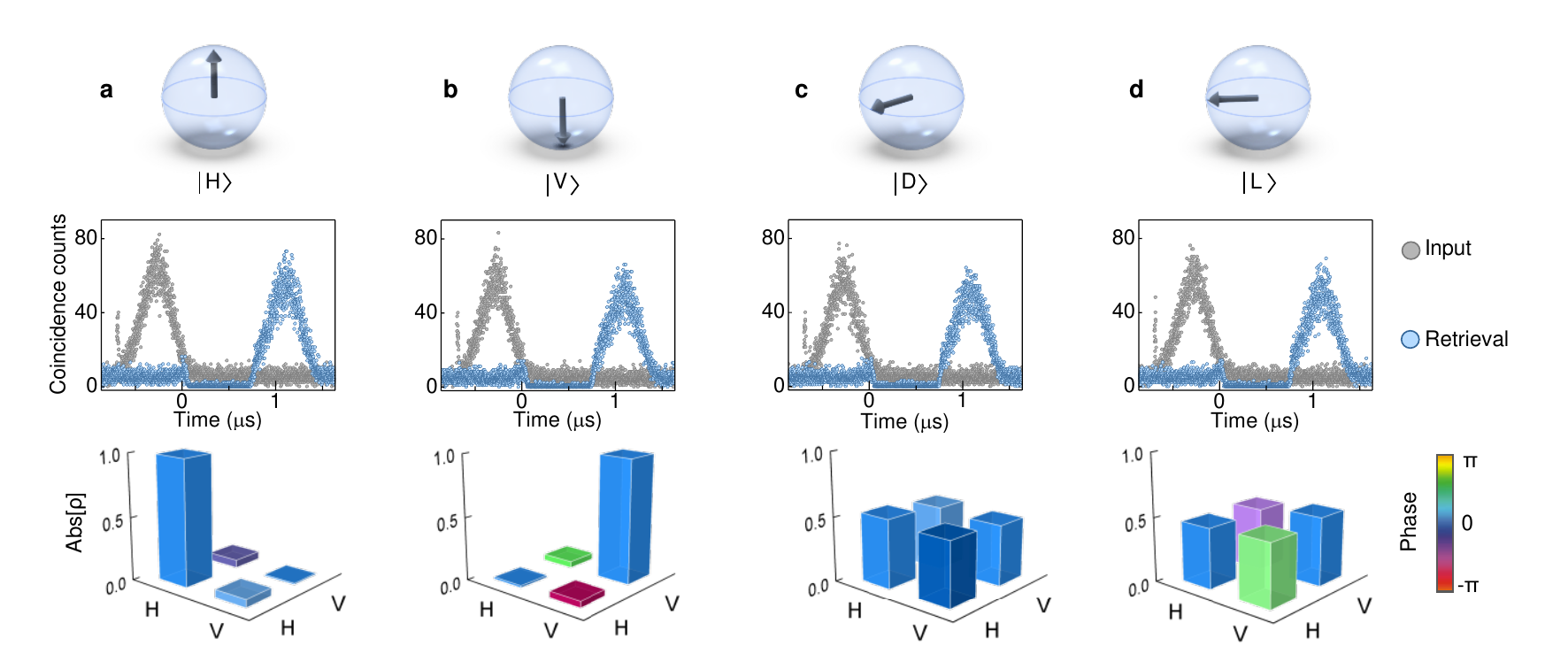}
\caption{\label{fig:qst}\textbf{Quantum memory for single-photon polarization qubits.}
\textbf{a,b,c, d}  show the quantum memory operations and performances for four choosen polarization qubit states: $|H\rangle$, $|V\rangle$, $|D\rangle=1/\sqrt{2}(|H\rangle+|V\rangle)$, and $|L\rangle=1/\sqrt{2}(|H\rangle+i|V\rangle)$. Each figure column presents the corresponding input polarization state on Bloch sphere, the temporal waveforms before and after storage, and the reconstructed density matrix of the retrieved qubits.}
\end{center}
\end{figure*}

\begin{figure*}
\begin{center}
\includegraphics[width=18cm]{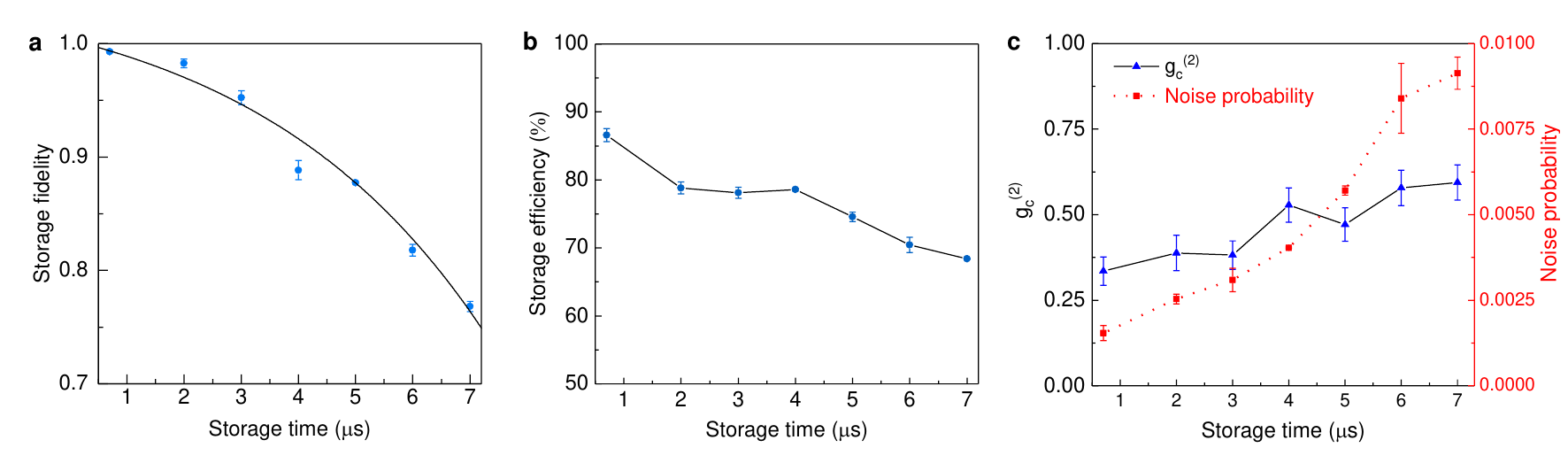}
\caption{\label{fig:fidelity}\textbf{The overall quantum memory performance}.  \textbf{a} the memory reterival qubit fidelity, \textbf{b} storage efficiency , and \textbf{c} $g_c^{(2)}$ and readout noise probability as functions of the storage time for the polarization qubit state $|L\rangle=1/\sqrt{2}(|H\rangle+i|V\rangle$. The error bars are from the standard deviation resulting from the statistical uncertainties of the measured coincidence counts.}
\end{center}
\end{figure*}

\bigskip
\noindent\textbf{Results}\\
\textbf{Double magneto-optical traps of cold atoms.}
The experimental setup, sketched in Fig.~\ref{fig:sys}(a) and further described in the Methods, consists of two cigar-shaped magneto-optical traps (MOT$_1$, MOT$_2$) of $^{85}$Rb atoms \cite{DMOTZhang}. MOT$_1$ is the heralded single photon source and MOT$_2$ is operated as the quantum memory. The whole experiment runs periodically with a repetition rate of 100~Hz. In each cycle of 10 ms, 9.7~ms MOT loading time is followed by 0.3 ms memory time window, as shown in Fig.~\ref{fig:sys}(b). In MOT$_1$, laser-cooled atoms are pumped to the lowest hyperfine level $|5S_{1/2}, F=2\rangle$, serving as a nonlinear medium for generating time-frequency entangled narrowband photon pairs using spontaneous four-wave mixing (SFWM) with counter-propagating pump and coupling laser beams \cite{DuOptica2014}. The atomic optical depth (OD) is 120 for MOT$_1$ on the transition $|5S_{1/2}, F=2\rangle\rightarrow|5P_{1/2}, F=3\rangle$. The MOT$_1$ magnetic field remains on all the time. In MOT$_2$, to minimize the decoherence effect induced by the inhomogeneous control light shifts due to the multi Zeeman substates and to avoid the off-resonant coupling between ground state and neighbouring excited state\cite{EIT2018Yu}, we optically pump the atoms to a specific Zeeman state $| 1\rangle=|5S_{1/2}, F=2, m_F=2\rangle$. The relevant atomic energy level diagram for the memory operation is shown in Fig.~\ref{fig:sys}(c). The circularly polarized memory control laser is on resonance to the transition $|2\rangle=|5S_{1/2},F=3, m_F=2\rangle\leftrightarrow|3\rangle=|5P_{1/2},F=3, m_F=3\rangle$. We set the control laser beam with an angle of $2.2^{o}$ to the memory modes, which is a key to suppress the scattering noise and FWM to the single-photon mode. Additional noise reduction is done though polarization  and frequency filtering. During the memory time, the MOT$_2$ magnetic field is switched off and the stray magnetic field is compensated by three pairs of Helmholtz coils down to 500 nT level. Because of the free expansion of the atoms, the OD for MOT$_2$ on the memory transition $|1\rangle\rightarrow|3\rangle$ varies linearly from 300 to 250 during the memory time. We vary the control laser power accordingly to maintain a constant EIT bandwidth, as shown Fig.~\ref{fig:sys}(b).

\bigskip
\noindent\textbf{Generation of heralded single photon polarization qubits with an optimal temporal waveform.}
A proper single photon source is necessary to match the spectral-temporal properties of the quantum memory for achieving a high storage efficiency \cite{QMemOptTheo2007, EIT2007Novikova, EIT2012Du}. Here, we implement the photon shaping technique by mapping the pump field spatial profile to the biphoton temporal waveform generated in MOT$_1$ \cite{sFWMSLM2015Du}. We focus the pump beam using a plano-convex lens to the center of the atomic cloud to generate the Gaussian-shaped biphoton waveform. As compared to other waveform shaping techniques using time-domain modulations \cite{HarrisPRL2008, DuPRL2010},  this method gives a high heralding efficiency and maintains the frequency entanglement between the paired Stokes and anti-Stokes photons. Under detection of a Stokes photon, its paired frequency anti-correlated anti-Stokes photon is projected to a pure single-photon state with well defined temporal waveform \cite{DuPRA2015, ChenPRL2016}:
\begin{equation}
\begin{split}\label{Eq:sFWM}
\psi_{in}(\tau) \sim f_{p}(L_1/2-V_{g}\tau)e^{-i\omega_{as}\tau},
\end{split}
\end{equation}
where $\tau$ is  the relative time respect to the detection of the Stokes photon, $L_1=1.7$ cm is the length of MOT$_1$, $V_{g}=2.0{\times}10^4{m/s}$ is the EIT group velocity of the anti-Stokes photon which can be controlled by the coupling laser intensity, and $\omega_{as}$ is the central angular frequency of the anti-Stokes photons. Fig.~\ref{fig:wfm2}(a) and (b) show the optimal Gaussian-shaped single-photon waveform with temporal length of about 700 ns (See Methods for details). The generated anti-Stokes  photons at MOT$_1$ are circularly polarized. After passing through a quart-wave plate, they are coupled into a polarization-maintaining single mode fiber. After the fiber output, an arbitrary polarization state $|S\rangle=cos\frac{\theta}{2}|H\rangle+e^{i\phi}sin\frac{\theta}{2}|V\rangle$ is created by passing the photon through a polarization qubit manipulation unit (QMU) with a combination of a quarter-wave plate (QWP) and a half-wave plate (HWP), where $\theta$ and $\phi$ are the angles denoting the polarization state on the Bloch sphere
presented in Fig.~\ref{fig:sys}(a). The entire heralded single photon polarization qubit is thus described by
\begin{equation}
\label{Eq:Qubit}
|\Phi\rangle=\psi_{in}(\tau)\otimes|S\rangle=\psi_{in}(\tau)\otimes[cos\frac{\theta}{2}|H\rangle+e^{i\phi}sin\frac{\theta}{2}|V\rangle].
\end{equation}
\\
\noindent\textbf{Quantum memory for single photon polarization qubits.} The EIT quantum memory is optimized  for one of the circular polarizations ($\sigma^+$) because of the atomic energy level configuration as shown in Fig.~\ref{fig:sys}(c). To store the polarization qubit state in the two-dimensional Hilbert space, we take the spatially-multiplexed dual-rail scheme \cite{LauratNC2018}  to convert the two polarization basis $|H\rangle$ and $|V\rangle$ into two spatial modes which can be stored simultaneously in the memory. As shown in Fig.~\ref{fig:sys}(a), the $|H\rangle$ and $|V\rangle$ polarization modes are spatially splitted after a beam displacer (BD). The $|V\rangle$ polarization is changed to $|H\rangle$ polarization after passing through a HWP. Then the both $|H\rangle$ polarized beams pass through a common QWP and become $\sigma^+$ circularly polarized for the memory. As the photon enters the memory, the control laser beam is switched off. After a programmable time delay, the control beam is switched on to read
out the photon spatial modes. Following a reversed space-to-polarization process, the two spatial modes are combined after the second BD. A polarization qubit analyzer consisting of a HWP, a QWP, and a PBS is used to measure the output photon polarization state.

To achieve  a high fidelity in polarization qubit memory, the two spatial channels CH$_h$ and CH$_v$ are required to have identical memory efficiency and performance. Under the optimal storage condition, for both channels CH$_h$ and CH$_v$,  we vary the control light Rabi frequency from $\Omega_c=10.2\gamma_{13}$ to $9.2\gamma_{13}$ linearly within the 0.3-ms memory time to compensate the time varying OD from 300 to 250. Both channels have the same dephasing rate $\gamma_{12}=0.0007\gamma_{13}$ between the states $|1\rangle$ and $|2\rangle$, with $\gamma_{13}=2\pi\times3$ MHz as the excited state dephasing rate resulting from the spontaneous emission. As presented in Fig.~\ref{fig:wfm2}(a) and (b), the retrieved temporal waveforms are nearly identical in both channels with a likeness as high as 96\%. The measured storage efficiencies at one-pulse delay are $(85.63\pm1.31)\%$ and $(86.31\pm1.32)\%$ for channels CH$_h$ and CH$_v$ respectively. The EIT transmission profiles are plotted in
Fig.~\ref{fig:wfm2}(c) for comparison. Figure \ref{fig:wfm2}(d) shows the measured storage efficiencies as functions of the storage time delay, displaying the same behaviour in both channels. The non-exponential decay of the storage efficiency results from the residual magnetic field gradient \cite{Kuzmich2005, EIT2018Yu}.

To confirm both memory channels preserve the single-photon quantum nature, we measure the conditional 2nd-order auto-correlation function $g_{c}^{(2)}$ of the retrieved photon (See Methods)\cite{HBTRef}. An ideal single photon Fock state gives $g_{c}^{(2)} = 0$ because a single photon cannot be divided into two and $g_{c}^{(2)}=0.5$ is for the two-photon state.  The measured $g_{c}^{(2)}$ over a coincidence window 700 ns as a function of the storage time are plotted in Fig.~\ref{fig:wfm2}(e), which shows  $g_{c}^{(2)}$ increases as the storage time becomes longer. The $g_{c}^{(2)}$ at zero storage time, as the reference value without storage, measures the quality of the heralded single photon source. For the storage time shorter than 3 $\mu s$, both channels are operated below the two-photon threshold with storage efficiencies above 80\%. The degraded $g_{c}^{(2)}$ after 3 $\mu s$ results from the noise photons and the drop of the storage efficiencies. Taking 3 $\mu s$ as our faithful memory time, we have a delay bandwidth product of 10, corresponding to a maximum 10 pulse delay for faithful single-photon storage.

After both channels are optimized for balanced performances, we start  the single-photon polarization qubit storage measurement to confirm the relative phase between the two channels is preserved during storage process. Our memory can be used to store an arbitrary unknown polarization qubit. As examples for illustration purpose, we perform storage for the following four states: $|H\rangle$, $|V\rangle$, $|D\rangle=1/\sqrt{2}(|H\rangle+|V\rangle)$, and $|L\rangle=1/\sqrt{2}(|H\rangle+i|V\rangle)$,  whose measured temporal waveforms and polarization density matrixes at one-pulse delay are depicted in Fig.~\ref{fig:qst}. The storage efficiencies and fidelities are presented in Table.~\ref{tbl:qst}. The qubit storage efficiency is above 85\% with a near-unity fidelity of 99\%. To show the overall memory performance, we plot the measured fidelity, storage efficiency, and $g_c^{(2)}$ for the state $|L\rangle$ as functions of the storage time in Fig.~\ref{fig:fidelity}. Within 3 $\mu s$, the high fidelity of $>$95\%, storage efficiency of $>(78.12\pm0.81)\%$, and $g_c^{(2)}<0.5$ are maintained. The fidelity displays a strong correlation with the storage efficiency and $g_c^{(2)}$. The fidelity degrades quickly after 3 $\mu s$ because of the reduction of the storage efficiency and increase of the two-photon probability.

\begin{figure*}
\begin{center}
\includegraphics[width=18cm]{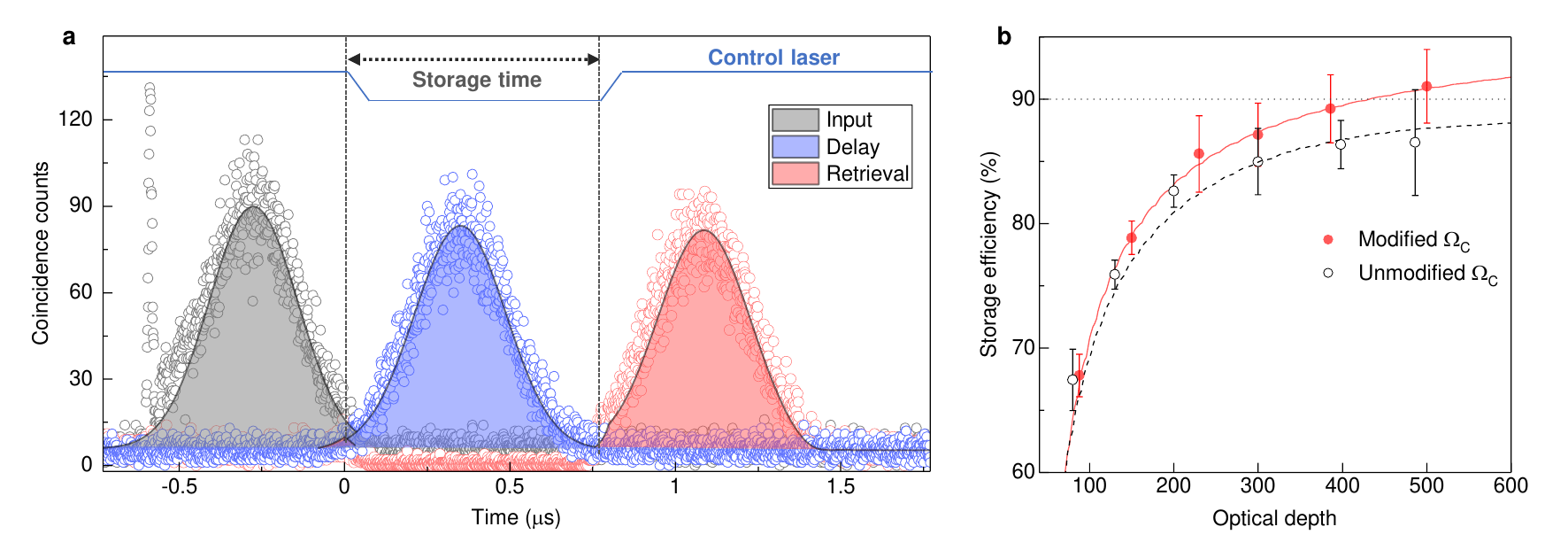}
\caption{\label{fig:wfm1}\textbf{Optimal storage of a single channel memory.} \textbf{a.} The input, EIT  delayed and retrieved temporal waveforms of the heralded single photons when the quantum momory is optimized for the horizontally polarized input optical channel $|H\rangle$. The photon coincidence counts are collected over 2400~s. \textbf{b.} The storage efficiency V.S. the optical depth of quantum memory. The solid lines are the best fitted theoretical wavform by fitting the input waveform using a Gaussian
function and then numerically calculated the retrieved waveform based on the measured experimental parameter of quantum memory. The red line denote the stituation when intensity of control light ($\Omega_c$) are modified to match the optical depth change during storage window and the black line denote the results when the intensity of control light are maintained constant.}
\end{center}
\end{figure*}

\begin{table}
\begin{tabular}{p{0.7cm}<{\centering}|p{2.4cm}<{\centering}|p{2.4cm}<{\centering}|p{2.4cm}<{\centering}}
\hline $|S\rangle$ & Storage Fidelity & Storage Efficiency & $g_c^{(2)}$ \\
\hline
$|H\rangle$ & $0.9988\pm0.0007$ & $(85.63\pm1.31)\%$ & $(0.342\pm0.031)$ \\
$|V\rangle$ & $0.9941\pm0.0016$ & $(86.31\pm1.32)\%$ & $(0.322\pm0.049)$ \\
$|D\rangle$ & $0.9969\pm0.0026$ & $(85.98\pm0.85)\%$ & $(0.338\pm0.063)$ \\
$|L\rangle$ & $0.9931\pm0.0045$ & $(86.59\pm0.95)\%$ & $(0.335\pm0.042)$ \\
\hline
\end{tabular}
\caption{\label{tbl:qst} \textbf{Quantum memory for single-photon polarization qubits.} The storage efficiencies and fidelities are measured at 1-bit (one-pulse) storage time for four different input polarization states. Error bars are $\pm$1 standard deviation.}
\end{table}

\bigskip
\noindent\textbf{Discussion}\\
Our efficient single-photon  polarization qubit quantum memory is realized on three key integrated elements in our system: cold atoms with high OD and low dephasing rate ($\gamma_{12}$), low noise, and single photons with controllable temporal waveform. The high OD of MOT$_2$ (250-300 in both channels) is achieved by implementing the dark-line MOT\cite{DMOTZhang} configuration with a large trapping volume (L$_2$=3 cm). The low ground-state dephasing rate is achieved by compensating the stray magnetic
field with three pairs of Helmholtz coils.

The noises in our measurements are mainly from the dark counts of SPCMs, the Rayleigh scattering and the spontaneous Raman scattering from the SFWM laser beams. The dark-count noise of our SPCMs is 25 counts per second, which can be further reduced by using better single-photon detectors. In the biphoton generation at MOT$_1$, the Rayleigh scattering noises from the pump and coupling beams are isolated from the Stokes and Anti-Stokes photon channels by using the FP etalon filters together with the spatial-mode filters. The spatial-mode filtering is achieved by setting the driving laser beams with angles to the single photon modes: In MOT$_1$ the Stokes/pump and anti-Stokes/coupling beams are aligned with an angle of 2.75$^o$, and in the quantum memory at MOT$_2$ the control beam is set with 2.2$^o$ to the single photon modes. Using single-mode fibers (SMFs) further reduces the Rayleigh scattering noises to an acceptable level. The polarization filter suppresses most of the Raman scattering in the heralded single photon channel, but the Raman photons with the same polarization are indistinguishable with the anti-Stokes photons and contribute to the major part of the noise. In the quantum memory side, we do not have observable Raman scattering from the control beam because of nearly zero population in the state $|2\rangle$.

Unlike other systems where FWM mixing nonlinear process may contribute noise photons to the memory and limits its efficiency \cite{EIT2014Buchler}, we have negligible FWM effect in our memory because of the phase mismatching from the large angle separation between the control light and the anti-Stokes photon modes. Meanwhile, FWM is much weaker in cold atoms compared to that in a hot vapor cell with the same OD \cite{EIT2018Yu,EIT2014Buchler}. As a result, the FWM noise is ignorable in our system. However, we observe a significant readout noise from the quantum memory depending on the residual vacuum pressure. A higher residual vacuum pressure gives higher readout noise rate. The readout noise photon counts keep increasing when storage time is increased, which attributes to the fast degradation of single photon nature and also the memory fidelity. Therefore, as shown in Fig.~\ref{fig:fidelity}(b), the 3 $\mu$s storage time is defined by the time when the retrieved photon losing its single-photon nature measured as $g^{(2)}_c$ $>$0.5. As a comparison, if the storage time is defined as the time when the efficiency drops by 50\%, our quantum memory has a storage time as long as 15 $\mu$s, which corresponding to a delay-bandwidth product of 50.


In our parameter region, we take the input state of $|L\rangle$ as an example, conditioned on a click of Stokes photon on the detector, the measured probabilities from the raw photon count data of retrieving a noise photon and retrieving a single photon are $0.15\%$ and $3.31\%$, $0.31\%$ and $3.39\%$ and $0.84\%$ and $2.95\%$ with a respective storage time of $0.7\mu s$, $3\mu s$ and $6\mu s$. Therefore, to extend the memory time, we can work with an ultra high vacuum chamber with lower vapor density of residual thermal atoms. Using single photon detectors with higher detection efficiency and lower dark counts also helps to extend the memory time to approach the storage time defined when the memory efficiency drops to 50\%.

In our system, the atomic OD slightly varies during the 0.3 ms storage time window because of the free falling and expansion of the cold atoms. Therefore, in order to achieve the best storage efficiency for the polarization qubit, we modify the coupling laser power to maintain a constant EIT bandwidth over the storage time windown, as depicted in Fig.~\ref{fig:sys}(b). This effect was investigated more detailly in the single channel storage, where the initial OD can go up to 500. As shown in Fig.~\ref{fig:wfm1}(b), comparing to using the constant control laser intensity, using the modified control laser shows significant improvement on the storage efficiency. Moreover, in the application where only one channel temporal mode storage is needed, we can optimize a single channel for a higher memory efficiency. Figure \ref{fig:wfm1}  shows the optimized single channel temporal waveform storage at one-pulse delay and we obtain the efficiency as high as 90.6\%. The measured dependence of storage efficiency on the initial $OD$ for this optimized single channel is shown in Fig.~\ref{fig:wfm1}(b).

In summary, we have demonstrated  an efficient EIT quantum memory for single-photon polarization qubits using laser cooled $^{85}$Rb atoms. With the balanced two-channel memory configuration, we have obtained the single-photon polarization qubit memory efficiency $>$85\% and fidelity $>$99\%. It is for the first time the achieved storage efficiency of any quantum memory for flying single-photon qubits exceeds 50\%, the threshold value for practical applications. Furthermore, our storage has the surpassed performance when comparing to the delay line based the optical fiber loop working on the same optical wavelength\cite{SE87GEM2016}. Thus  the quantum memory developed here may become a critical component of a quantum network and quantum computer. Looking forward, while the quantum memory efficiency has been approaching unit, its capacity and scalability is still need to be improved. Meanwhile a practical integrated quantum application system requires further innovations in production of cold atoms with a larger duty cycle, generation of narrowband single photons with high purity, and single-photon detection with higher efficiency and lower dark counts.

\bigskip
\noindent\textbf{Methods}\\
{\footnotesize
\noindent\textbf{Experimental setup.}
The experimental setup consists of two two-dimensional dark-line MOTs (MOT$_1$, MOT$_2$) of $^{85}$Rb atoms \cite{DMOTZhang}. The heralded single photons generated from MOT$_1$ are collected with a polarization-maintaining single mode fibre, and then transmitted 15 meters away to MOT$_2$ for storage that is built on another optical table. For MOT$_1$, the typical OD is 120. The gradient of the magnetic field is about 8 G/cm. Each trapping laser beam has a power of 15 mW and a diameter of 2 cm. The total power of the two repump laser beams is 25 mW with the same diameter as the trapping laser beam.  Two dark lines with 0.75 mm and 1.0 mm diameter, respectively, are positioned in the center region of the two repump laser beams. Two 4-f imaging setups are used to project the dark lines onto the atomic ensemble.

For MOT$_2$,  the typical OD is 500 for single path setup and 300 for the dual rail setup. The magnetic field gradient is about 8 G/cm. Each trapping laser beam has a power of 50 mW and a diameter of 3.3 cm.  The total power of the two repump laser beams is 40 mW with the same diameter as the trapping beam. Two dark lines with 0.75 mm and 1.5 mm width, respectively, are positioned in the center region of the two repump laser beams.  Two 4-f imaging setups are applied to project the dark lines onto the atomic ensemble.

\bigskip
\noindent\textbf{Heralded single photons.}
The heralded single photons are obtained from a time-frequency entangled photon pair (biphoton) source using SFWM \cite{DuOptica2014} at MOT$_1$. In presence of counter-propagating circularly polarized pump (780 nm, polarization $\sigma^{-}$, 10 $\mu$W) and coupling (795 nm, $\sigma^{+}$, 3.5 mW) laser beams, phase-matched Stokes (780 nm, $\sigma^{-}$) and anti-Stokes (795 nm, $\sigma^{+}$) photons are generated in opposing directions, as illustrated in Fig.~\ref{fig:sys}(a). The pump laser  is blue detuned from the transition $|5S_{1/2},F=2\rangle\leftrightarrow|5P_{3/2},F=3\rangle$ by 80~MHz. The pump beam is focused on the center of MOT$_1$ with a  0.82~mm $1/e^2$ diameter. The coupling laser,  resonant to the transition $|5S_{1/2},F=3\rangle\leftrightarrow|5P_{1/2},F=3\rangle$, is a collimated Gaussian beam with a $1/e^2$ diameter of 1.8 mm. The angle between the Stokes/anti-Stokes axis and the pump/coupling axis is chosen to be $2.75^{o}$, which provides spatial separation of the weak Stokes and anti-Stokes fields from the strong pump and coupling fields. The paired photons are collected into two SMFs whose spatial modes are mirrored with each other and have overlapped focus at the MOT$_1$ center with a $1/e^2$ diameter of 0.24~mm. The biphoton coherence time is about 700 ns, which corresponding to a bandwidth of 0.39 MHz. Two Fabry-Perot (FP) filters that are not shown in Fig.~\ref{fig:sys}(a) with bandwidth of 210 MHz and isolation ratio of 42~dB are used to further reduce stray noise photons in both Stokes and anti-Stokes modes. Two single-photon counting modules (SPCM, Perkin Elmer, SPCM-AQRH-16) is used to detect both Stokes and anti-Stokes photons. The Stokes and anti-Stokes coincidence counts are recorded by a time-to-digital converter (Fast Comtec P7888) with a time bin of 1~ns. The measured biphoton count rate is 13 pairs/s. Taking into account the SMF fiber-fiber coupling efficiency (80\%), filter transmission (60\%), SPCM detection efficiency (60\%), memory channel collection efficiency (72\%), and duty cycle (3\%), we estimate the photon pair generation rate from the source to be about 5800 pairs/s. Under detection of a Stokes photon, the heralded anti-Stokes photon is projected into a pure single photon state with a well-defined temporal waveform \cite{DuPRA2015,ChenPRL2016}. The experimentally measured heralding efficiency is 3.2\%, corresponding to 15.5\% heralding efficiency from the source with all the losses corrected. These losses increase the measurement time, but have no effect on the quantum memory performance.

The optimal Gaussian-like single-photon temporal waveform is obtained  following the optimization procedure described in Refs\cite{EIT2007Novikova, EIT2012Du}.

\bigskip
\noindent\textbf{Quantum memory setup.}
The cold $^{85}$Rb medium at MOT$_2$ serves as the quantum memory with the size of 4 $\times$ 4 $\times$ 30 mm$^{3}$. For the two-channel qubit quantum memory, their balanced OD is 300, while for a single channel the optimal OD can be as high as 500. The temperature of the atomic cloud was about 200 $\mu$K. The signal beam is focused at the center of MOT$_2$ with a beam waist radius of 150 $\mu$m .  In each operation cycle, the MOT$_2$ quadruple gradient magnetic field is switched off before the 0.3 ms memory window. At beginning of the memory time window, a collimated control laser beam with a 4.7~mm $1/e^2$ diameter is switched on. After a Stokes photon is detected and the heralded anti-Stokes photon enters inside the atomic medium, we switch off the control laser with a falling time of 70 ns using an acousto-optic modulator (AOM) driven by a digital waveform generator (Rigol DSG815), and convert the flying anti-Stokes photonic state into the longly-lived atomic state. After a controllable time delay, we switch on the control laser to retrieve the photon state.

\bigskip
\noindent\textbf{Storage efficiency.}
The storage efficiency of the quantum memory is defined as the probability of storing and retrieving a  single photon:
\begin{eqnarray}
\begin{array}{ll}
&\eta=\frac{\int |\psi_{out}(\tau)|^{2} d\tau}{\int |\psi_{in}(\tau)|^{2}d\tau}
\end{array}
\end{eqnarray}
where $\psi_{in}(\tau)$ and $\psi_{out}(\tau)$ are the unnormalized input and output heralded single-photon wavepackets. Heralded by the detection of Stokes photon, $\int |\psi_{in}(\tau)|^{2}d\tau$ ($\int |\psi_{out}(\tau)|^{2} d\tau$) represent the probability of sending (retrieving) a single photon into (from) the memory. In our experiment, the integration window is chosen to be symmetric around the peak of Gaussian shape single photon waveform with a width of 700 ns, which is wide enough to contain the entire photon waveform. We notice that there exists a sharp spike in front of the main Gaussian wavepacket of the input heralded single photon. This sharp spike, contributing around 3\% of the photon counts, is the optical precursor of the single photon that does not interact with the atoms \cite{SPhPrecursor,BPhPrecursor} . Therefore we exclude the optical precursor for counting the memory efficiency in our work.

We note that the measured storage efficiency is an averaged value in the 0.3 ms duty time window on which the atomic OD decreases over time. If we have single photons on demand, we can prepare the cold atoms with the highest OD just before the photons coming and the memory efficiency can be much higher.

\bigskip
\noindent\textbf{Conditional 2nd order auto-correlation function.}
The quantum particle nature of the heralded anti-Stokes photon is verified by measuring its conditional 2nd order auto-correlation function with a Hanbury, Brown, and Twiss interferometer. Under the condition of detecting a Stokes photon, we send the heralded anti-Stokes photon to a 50/50 fiber beam splitter. A measure of the quality of heralded single photon is given by the conditional 2nd order auto-correlation function \cite{HBTRef} :
\begin{eqnarray}
\begin{array}{ll}
&g_{c}^{(2)}=\frac{N_{GTR}N_{G}}{N_{GT}N_{GR}}
\end{array}
\end{eqnarray}
where $N_{G}$ is the Stokes photon counts, $N_{GR}$($N_{GT}$) is the two-fold coincidence counts between the Stokes photon detector and the reflection (transmission) output port of the FBS, and $N_{GTR}$ is the threefold coincidence counts. The coincidence counts are measured over a time window of 700 ns.

\bigskip
\noindent\textbf{Polarization qubit memory fidelity.}
The fidelity of qubit storage-readout operation is defined as:
 \begin{eqnarray}
\begin{array}{ll}
&F=\lvert Tr(\sqrt{\sqrt{\rho_{in}} \rho_{out} \sqrt{\rho_{in}}} )\rvert ^2
\end{array}
\end{eqnarray}
where $\rho_{in}$ and $\rho_{out}$ are the polarization density matrix of input and retrieved single photon qubit, respectively.

\bigskip
\noindent\textbf{Waveform likeness.}
The temporal waveform likeness between the reterived and input single photon is defined as:
\begin{eqnarray}
\begin{array}{ll}
&\frac{|\int \psi_{in}(\tau)\psi_{out}(\tau)d\tau|^2}{\int |\psi_{in}(\tau)|^2 d\tau \int |\psi_{out}(\tau)|^2 d\tau}
\end{array}
\end{eqnarray}

\bigskip
\noindent\textbf{Theoretical simulation of the quantum storage}
A simple theoretical model based on the Maxwell-Schr\"{o}dinger equation of single photon field and the optical  Bloch equation of atomic ensembles are implemented to describe our memory dynamics:
\begin{eqnarray}
\begin{array}{ll}
&(\partial_{\tau}{} + c_0\partial_{\tilde{z}}{})\tilde{\varepsilon}_{as}= ig\sqrt{N}\tilde{P} \\
&\partial_{\tau}{\tilde{P}}= -\gamma_{13}\tilde{P}+ \frac{ig\sqrt{N}}{2}\tilde{\varepsilon}_{as}+\frac{i}{2}\tilde{\Omega}_{c2}\tilde{S} \\
&\partial_{\tau}{\tilde{S}}= -\gamma_{12}\tilde{S}+\frac{i}{2}\tilde{\Omega}^*_{c2}\tilde{P}
\end{array}
\end{eqnarray}
where $\tilde{\varepsilon}_{as}$ is the time- and position-dependent slow  varying envelop of the heralded single photon's quantum field and $\tilde{\Omega}_{c2}$ is the Rabi frequency of the control laser. $\tilde{P}$ and $\tilde{S}$ are the slow varying envelope of the collective polarization of $|1\rangle\leftrightarrow|3\rangle$ coherence and
$|1\rangle\leftrightarrow|2\rangle$ coherence, respectively. $g$ is the photon-atom coupling strength and can be obtained from the relation $OD=g^2NL_2/(\gamma_{13}c_0)$. $N$ is the total atom number in the interaction volume with the assumption that atoms are distributed uniformly over MOT$_2$. $L_2$ is the MOT$_2$ length. $c_0$ is the optical speed in vacuum. With the anti-Stokes entrance position as $\tilde{z}=0$ in MOT$_2$, we thus have $\psi_{in}(\tau)=\psi_{as}(\tau)$ and $\tilde{\varepsilon}_{as}(\tau,\tilde{z}=0)=\psi_{as}(\tau)e^{i\omega_{as}\tau}$. After numerically solving Eq. (3) we obtain $\tilde{\varepsilon}_{as}(\tau,L_2)$ and then $\psi_{out}(\tau)$  can be derived from the relation $\psi_{out}(\tau)=\tilde{\varepsilon}_{as}(\tau,L_2)e^{-i\omega_{as}\tau}$.
}


\bigskip
\bigskip\noindent\textbf{Acknowledgements}\\
\noindent This work was supported by the NKRDP of China (Grants No. 2016YFA0301803 and No. 2016YFA0302800), the NSF of  China (Grants No. 11474107, No. 61378012, No. 91636218, No. 11822403, No. 11804105), the GDSBPYSTIT (Grant No.2015TQ01X715), the GNSFDYS (Grant No. 2014A030306012). S.D. acknowledges the support from Hong Kong Research Grants Council (Project Nos. 16304817 and C6005-17G).

\bigskip\noindent\textbf{Author Contributions}\\
\noindent S.C.Z., S.D., H.Y. and S.L.Z. designed the experiment.
Y.F.W., J.F.L., S.C.Z., K.Y.S., Y.R.Z. and K.Y.L. carried out the experiments.
Y.F.W., J.F.L., S.C.Z., K.Y.S. and H.Y. conducted raw data analysis.
S.C.Z., S.D., H.Y. and S.L.Z. wrote the paper, and all authors discussed the paper contents.
H.Y. and S.L.Z. supervised the project.

\bigskip
\noindent\textbf{Competing Financial Interests}\\
\noindent The authors declare no competing financial interests.
\end{document}